\documentclass[prb,aps,showpacs,onecolumn]{revtex4}
\usepackage{amssymb}
\usepackage{amsfonts}
\usepackage{amsmath}
\usepackage{graphicx}
\usepackage{bm}

\begin{document}

\title{Solvable rational extensions of the isotonic oscillator}
\author{Yves Grandati }
\affiliation{Institut de Physique, Equipe BioPhyStat, ICPMB, IF CNRS 2843, Universit\'{e}
Paul Verlaine-Metz, 1 Bd Arago, 57078 Metz, Cedex 3, France}

\begin{abstract}
\bigskip Combining recent results on rational solutions of the Riccati-Schr%
\"{o}dinger equations for shape invariant potentials to the finite
difference B\"{a}cklund algorithm and specific symmetries of the isotonic
potential, we show that it is possible to generate the three infinite sets ($%
L1$, $L2$ and $L3$ families) of regular rational solvable extensions of this
potential in a very direct and transparent way.
\end{abstract}

\maketitle

\bigskip


\section{Introduction}

In quantum mechanics, the rarity of the potentials which are exactly
solvable in closed-form (most of them belonging to the class of
shape-invariant potentials \cite{cooper,Dutt,Gendenshtein}) gives a
undeniable importance to the reseach of new families of such potentials. A
possible way to generate new solvable potentials is to start from the known
ones and to construct regular rational extensions of them. If the procedure
has a long history, in the last years important progress have been made in
this direction \cite%
{gomez,gomez2,gomez3,gomez4,gomez5,quesne1,quesne,quesne2,quesne3,odake,sasaki,ho,odake2,sasaki2,dutta}%
. In a recent work \cite{grandati2} we proposed an approach allowing to
generate such regular extensions starting from every translationally
shape-invariant potential (TSIP) of the second category (as defined in \cite%
{grandati}). For this, we use regularized excited states Riccati-Schr\"{o}%
dinger (RS) functions as superpotentials in a generalized SUSY partnership.
The regularization scheme corresponds to a "spatial Wick rotation" which
eliminates the singularities from the real axis, a device already suggested
by Shnol' \cite{shnol'} in 1994 as a way to generate rational extensions of
the harmonic potential. In the following years, this suggestion has been
developped by Samsonov and Ovcharov \cite{samsonov} and Tkachuk \cite%
{tkachuk}. Recently Fellows and Smith \cite{fellows} rediscovered this
technique in the case of the harmonic oscillator, the second rational
extension of which being the so-called CPRS potential \cite{carinena}. In 
\cite{grandati2}, we have extended the procedure to cover the whole set of
TSIP belonging to the second category. For the isotonic oscillator, we
recovered the $L1$ family of rational extensions discovered by Gomez-Ullate,
Kamran and Milson \cite{gomez,gomez2,gomez3,gomez4,gomez5}, Quesne \cite%
{quesne1,quesne,quesne2,quesne3} and Odake, Sasaki et al \cite%
{odake,odake2,sasaki2}. For the other second category potentials, the
infinite set of regular quasi-rational extensions that we obtain coincides
with the $J1$ family \cite%
{gomez,gomez2,gomez3,gomez4,quesne1,quesne,quesne2,quesne3,odake,odake2,sasaki2}%
.

In the present article, combining the finite difference B\"{a}cklund
algorithm with new regularization schemes which are based on specific
symmetries of the isotonic potential, we show how the extension of the SUSY
QM partnership to excited states allows to generate the three infinite sets $%
L1$, $L2$ and $L3$ of regular rationally solvable extensions of the isotonic
potential (as well as the singular $L0$ and $L3$ ones) in a direct and
systematic way. This approach leads to a simple and transparent proof of the
shape-invariance of the potentials of the $L1$ and $L2$ series.

The paper is organized as follows. We first recall how the generalization of
the SUSY partnership based on excited states leads to a series of singular
rational extensions of the initial potential. We then introduce basic
elements concerning the finite difference B\"{a}cklund algorithm viewed as a
set of covariance transformations for the class of Riccati-Schr\"{o}dinger
equations and we interpret the generalized SUSY partnership in this
perspective. In the third and fourth sections, we recapitulate some results
concerning the isotonic oscillator, its connection with confluent
hypergeometric equation and the Kienast-Lawton-Hahn's Theorem which
describes the distribution of the zeros of the Laguerre functions on the
real axis. The fifth section is devoted to present the set of parameters
transforms which are discrete symmetries of the isotonic potential. Using
them as regularization transformations, we show then that the finite
difference B\"{a}cklund algorithm based on the corresponding regularized RS
functions generates directly the three series $L1$, $L2$ and $L3$ of regular
rationally solvable extensions of the isotonic potential. In the last
section, we prove the shape-invariance of the potentials of the $L1$ and $L2$
series.

\section{Generalized SUSY partnership based on excited states: $L0$ series
of rational extensions}

Consider a family of closed form exactly solvable hamiltonians $%
H(a)=-d^{2}/dx^{2}+V(x;a),\ a\in \mathbb{R}^{m},\ x\in I\subset \mathbb{R}$,
the associated bound states spectrum of which being given by $\left(
E_{n}(a),w_{n}(x;a\right) )$, where $w_{n}(x;a)=-\psi _{n}^{\prime
}(x;a)/\psi _{n}(x;a)$ is the Riccati-Schr\"{o}dinger (RS) function
associated to the $n^{th}$ bound state eigenfunction $\psi _{n}(x;a)$. The
Riccati-Schr\"{o}dinger (RS) equation \cite{grandati} for the level $%
E_{n}(a) $ is then

\begin{equation}
-w_{n}^{\prime }(x;a)+w_{n}^{2}(x;a)=V(x;a)-E_{n}(a),  \label{edr4}
\end{equation}%
where we suppose $E_{0}(a)=0$. The RS function presents $n$ real
singularities associated to the $n$ simple nodes of the eigenstates $\psi
_{n}(x;a)$. As it is well known \cite{robnik,klippert}, $H(a)$ admits
infinitely many different factorizations of the form

\begin{equation}
H(a)-E_{n}(a)=A^{+}\left( w_{n}\right) A\left( w_{n}\right) ,
\end{equation}%
where

\begin{equation}
A\left( w_{n}\right) =d/dx+w_{n}(x;a),  \label{opA}
\end{equation}%
with, in particular

\begin{equation}
A\left( w_{n}\right) \psi _{n}(x;a)=0.
\end{equation}

This allows to associate to $H(a)$ or $V(x;a)$ an infinite family of
partners given by

\begin{equation}
H^{\left( n\right) }(a)-E_{n}(a)=A\left( w_{n}\right) A^{+}\left(
w_{n}\right) =-d^{2}/dx^{2}+V^{\left( n\right) }(x;a),
\end{equation}%
with

\begin{equation}
V^{\left( n\right) }(x;a)=V(x;a)+2w_{n}^{\prime }(x;a).
\end{equation}

For $n\geq 1$, these potentials are all singular at the nodes of $\psi
_{n}(x;a)$ and are defined on open intervals only.

On these domains, $H^{\left( n\right) }(a)$ is (quasi)isospectral to $H(a)$.
Indeed, writing\bigskip 
\begin{equation}
\psi _{k}^{\left( n\right) }(x;a)=A\left( w_{n}\right) \psi _{k}(x;a),
\label{fopartner}
\end{equation}%
it is easy to verify that we have for any $k$

\begin{equation}
H^{\left( n\right) }(a)\psi _{k}^{\left( n\right) }(x;a)=E_{k}(a)\psi
_{k}^{\left( n\right) }(x;a),  \label{hampart}
\end{equation}%
that is, $\psi _{k}^{\left( n\right) }(x;a)$ is an eigenstate of $H^{\left(
n\right) }(a)$ associated to the eigenvalue $E_{k}(a)$. We write symbolically

\begin{equation}
V^{\left( n\right) }(x;a)\underset{iso}{\equiv }V(x;a),
\end{equation}%
where $\underset{iso}{\equiv }$ means "isospectral to". Defining

\begin{equation}
w_{n,k}(x;a)=-\frac{\psi _{k}^{\left( n\right) \prime }(x;a)}{\psi
_{k}^{\left( n\right) }(x;a)},  \label{RSpart1}
\end{equation}%
Eq.(\ref{hampart}) gives, for $k>n$

\begin{equation}
-w_{n,k}^{\prime }(x;a)+w_{n,k}(x;a)^{2}=V^{\left( n\right) }(x;a)-E_{k}(a).
\end{equation}

This scheme generalizes the SUSY QM partnership, by using the excited state
RS functions $w_{n}$ as superpotentials. However, only for the ground state $%
n=0$, the factorization and then the partner potential $V^{\left( 0\right)
}(a)=V(x;a)+2w_{0}^{\prime }(x;a)$ are non singular and we recover the usual
SUSY QM partnership \cite{cooper,Dutt}.

\section{Finite difference B\"{a}cklund algorithm}

We can consider the preceding partnership in a different way which gives a
prominent role to the covariance transform of the RS equations class.

\subsection{Invariance group of the Riccati equations}

As established by Cari\~{n}ena et al. \cite{carinena2,Ramos}, the \bigskip
finite-difference B\"{a}cklund algorithm is a consequence of the invariance
of the set of Riccati equations under a subset of the group $\mathcal{G}$ of
smooth $SL(2,\mathbb{R})$-valued curves $Map(\mathbb{R},SL(2,\mathbb{R}))$.
For any element $A\in $ $\mathcal{G}$ characterized by the matrix:

\begin{equation}
A(x)=\left( 
\begin{array}{cc}
\alpha (x) & \beta (x) \\ 
\gamma (x) & \delta (x)%
\end{array}%
\right) ,\quad \det A(x)=\alpha (x)\delta (x)-\beta (x)\gamma (x)=1,
\label{matrice}
\end{equation}%
the action of $A$ on $Map(\mathbb{R},\overline{\mathbb{R}})$ is given by:

\begin{equation}
w(x)\overset{A}{\rightarrow }\widetilde{w}(x)=\frac{\alpha (x)w(x)+\beta (x)%
}{\gamma (x)w(x)+\delta (x)}.  \label{transfo}
\end{equation}

If $A$ acts on a solution of the Riccati equation:

\begin{equation}
w^{\prime }(x)=a_{0}(x)+a_{1}(x)w(x)+a_{2}(x)w^{2}(x),  \label{edrg}
\end{equation}%
we obtain a solution of a new Riccati equation:

\begin{equation}
\widetilde{w}^{\prime }(x)=\widetilde{a}_{0}(x)+\widetilde{a}_{1}(x)%
\widetilde{w}(x)+\widetilde{a}_{2}(x)\widetilde{w}^{2}(x),  \label{edr2}
\end{equation}%
the coefficients of which being given by

\begin{equation}
\overrightarrow{\widetilde{a}}(x)=M(A)\overrightarrow{a}(x)+\overrightarrow{W%
}(x),\quad \overrightarrow{u}(x)=\left( 
\begin{array}{c}
u_{2}(x) \\ 
u_{1}(x) \\ 
u_{0}(x)%
\end{array}%
\right) ,  \label{transfocoeff1}
\end{equation}%
where:

\begin{equation}
M(A)=\left( 
\begin{array}{ccc}
\delta ^{2}(x) & -\gamma (x)\delta (x) & \gamma ^{2}(x) \\ 
-2\beta (x)\delta (x) & \alpha (x)\delta (x)+\beta (x)\gamma (x) & -2\alpha
(x)\gamma (x) \\ 
\beta ^{2}(x) & -\alpha (x)\beta (x) & \alpha ^{2}(x)%
\end{array}%
\right) ,\quad \overrightarrow{W}(x)=\left( 
\begin{array}{c}
W(\gamma ,\delta ;x) \\ 
W(\delta ,\alpha ;x)+W(\beta ,\gamma ;x) \\ 
W(\alpha ,\beta ;x)%
\end{array}%
\right)
\end{equation}%
$(W(f,g;x)=f(x)g^{\prime }(x)-f^{\prime }(x)g(x)$ is the wronskian of $f(x)$
and $g(x)$ in $x$). As noted in \cite{carinena2}, Eq.(\ref{transfocoeff1})
defines an affine action of $\mathcal{G}$ on the set of general Riccati
equations.

\subsection{Particular case of the RS equations and finite difference B\"{a}%
cklund algorithm}

The most general elements of $\mathcal{G}$ preserving the subset of RS
equations has been determined in \cite{carinena2}. Among them we find in
particular the elements of the form:

\begin{equation}
A(\phi )=\frac{1}{\sqrt{\lambda }}\left( 
\begin{array}{cc}
\phi (x) & \lambda -\phi ^{2}(x) \\ 
-1 & \phi (x)%
\end{array}%
\right) ,\ \lambda >0,  \label{transfo2}
\end{equation}%
where $\phi (x)$ satisfies an RS equation with the same potential as in Eq.(%
\ref{edr4}) but with a shifted energy:

\begin{equation}
-\phi ^{\prime }(x)+\phi ^{2}(x)=V(x)-\left( E-\lambda \right) .
\end{equation}

With this choice $\widetilde{w}(x)$ satisfies the RS equation:

\begin{equation}
-\widetilde{w}^{\prime }(x)+\widetilde{w}^{2}(x)=\widetilde{V}_{\phi
}(x)-\lambda ,
\end{equation}%
where $\widetilde{V}_{\phi }(x)=V(x)+2\phi ^{\prime }(x)$.

Consequently, starting from a given RS function of the discrete spectrum $%
w_{n}(x;a)$, for every value of $k$ such that $E_{k}>E_{n}$ , we can build
an element $A\left( w_{n}\right) \in \mathcal{G}$ of the form:

\begin{equation}
A\left( w_{n}\right) =\frac{1}{\sqrt{E_{k}(a)-E_{n}(a)}}\left( 
\begin{array}{cc}
w_{n}(x;a) & E_{k}(a)-E_{n}(a)-w_{n}{}^{2}(x;a) \\ 
-1 & w_{n}(x;a)%
\end{array}%
\right)  \label{transfoback}
\end{equation}%
which transforms $w_{k}$ as:

\begin{equation}
w_{k}(x;a)\overset{A\left( w_{n}\right) }{\rightarrow }w_{k}^{\left(
n\right) }(x;a)=-w_{n}(x;a)+\frac{E_{k}(a)-E_{n}(a)}{w_{n}(x;a)-w_{k}(x;a)},
\label{transfoback2}
\end{equation}%
where $w_{k}^{\left( n\right) }$ is a solution of the RS equation:

\begin{equation}
-w_{k}^{\left( n\right) \prime }(x;a)+\left( w_{k}^{(n)}(x;a)\right)
^{2}=V^{\left( n\right) }(x;a)-E_{k}(a),  \label{eqtransform}
\end{equation}%
with the same energy $E_{l}(a)$ as in Eq(\ref{edr4}) but with a modified
potential

\begin{equation}
V^{\left( n\right) }(x;a)=V(x;a)+2w_{n}^{\prime }(x;a).
\end{equation}

This is the content of the finite-difference B\"{a}cklund algorithm \cite%
{carinena2,Ramos,Fernandez,Mielnik,Adler1,Adler2}. It transposes at the
level of the RS equations the covariance of the set of Schr\"{o}dinger
equations under Darboux transformations \cite{darboux,luban,matveev}. In the
following we call $A\left( w_{n}\right) $ a Darboux-B\"{a}cklund
Transformation (DBT).

To $V(x;a)$, $A\left( w_{n}\right) $ associates the (quasi)isospectral
partner $V^{\left( n\right) }(x;a)$. Among the $A\left( w_{n}\right) $, only 
$A\left( w_{0}\right) $ leads to the regular, usual SUSY QM partner $%
V^{\left( 0\right) }(x;a)$. The correspondence between the eigenvalues of $%
V(x;a)$ and $V^{\left( n\right) }(x;a)$ is direct. We also have from Eq(\ref%
{fopartner}) and Eq(\ref{opA})

\begin{equation}
\psi _{k}^{\left( n\right) }(x;a)\sim \left( w_{n}(x;a)-w_{k}(x;a)\right)
\psi _{k}(x;a),
\end{equation}%
that is (see Eq(\ref{RSpart1})),

\begin{equation}
w_{n,k}(x;a)=w_{k}^{\left( n\right) }(x;a).
\end{equation}

Then, the finite difference B\"{a}cklund algorithm generates exactly the RS
functions corresponding to the spectrum of the generalized SUSY partner $%
V^{\left( n\right) }$ of $V$.

Note that for shape invariant potentials (SIP) \cite%
{cooper,Dutt,Gendenshtein}, $A\left( w_{0}\right) $ is in fact an invariance
transformation of the RS equations associated to the considered family of
potentials (indexed by the multiparameter $a$), since in this case

\begin{equation}
V^{\left( 0\right) }(x;a)=V(x;a_{1})+R(a)
\end{equation}%
and 
\begin{equation}
w_{k}^{\left( 0\right) }(x,a)=w_{k-1}(x,a_{1}),
\end{equation}%
where $a_{1}=f(a)$ and $R(a)$ are two given functions of the multiparameter $%
a$.

As we noted before, starting from the RS function $w_{n}$ of a regular
excited bound state which has $n$ nodes on the real domain of definition $I$
of $V(x;a)$, we generate via $A\left( w_{n}\right) $ a generalized SUSY
partner which presents $n$ singularities on this domain. Nevertheless, the
finite difference B\"{a}cklund algorithm can be applied by replacing $w_{n}$
by any other solution of the same RS equation Eq(\ref{edr4}), even if this
solution does not correspond to a physical state. Knowing $w_{n}(x,a)$, the
general solution of Eq(\ref{edr4}) is given by

\begin{equation}
W_{n}(x;a,W_{0})=w_{n}(x;a)-\frac{e^{2\int_{x_{0}}^{x}w_{n}(s;a)ds}}{%
W_{0}+\int_{x_{0}}^{x}dse^{2\int_{x_{0}}^{s}w_{n}(t;a)dt}},  \label{RSgensol}
\end{equation}%
where $W_{0}$ is an arbitrary real parameter. We could then use the DBT $%
A\left( W_{n}\right) $ to build a generalized SUSY partner potential $%
V^{\left( n\right) }(x;a,W_{0})=V(x;a)+2W_{n}^{\prime }(x;a,W_{0})$ and look
for values of $W_{0}$ for which $W_{n}$ and $V^{\left( n\right) }$ are not
singular. For some potentials it is nevertheless possible, by using specific
symmetries, to build directly the researched regular RS\ functions. Such
symmetries exist in particular for the isotonic oscillator.

\section{The isotonic oscillator}

As shown in \cite{grandati}, the primary translationally shape invariant
potentials (TSIP), for which $a_{1}=a+\alpha $, can be classified into two
categories in which the potential can be brought into a harmonic or isotonic
form respectively, using a change of variables which satisfies a constant
coefficient Riccati equation.

The first element of the second category is the isotonic oscillator
potential itself (ie the radial effective potential for a three dimensional
isotropic harmonic oscillator with zero ground-state energy) defined on the
positive real half line

\begin{equation}
V(x;\omega ,a)=\frac{\omega ^{2}}{4}x^{2}+\frac{a(a-1)}{x^{2}}+V_{0}(\omega
,a),\ x>0,  \label{isotpot}
\end{equation}%
with $a=l+1\geq 1$ and $V_{0}(\omega ,a)=-\omega \left( a+\frac{1}{2}\right) 
$. The shape invariance property of $V(x;\omega ,a)$ is expressible as

\begin{equation}
V^{\left( 0\right) }(x;\omega ,a)=V(x;\omega ,a_{1})+2\omega  \label{VSIP1}
\end{equation}%
and its spectrum is given by

\begin{equation}
E_{n}\left( \omega \right) =2n\omega ,\ \psi _{n}\left( x;\omega ,a\right)
\sim \exp \left( -\int w_{n}\left( x;\omega ,a\right) dx\right) ,
\label{spectrisot}
\end{equation}%
where the excited state Riccati-Schr\"{o}dinger function (RS function) $%
w_{n}\left( x;\omega ,a\right) $ can be written as a terminating continued
fraction as

\begin{equation}
w_{n}(x;\omega ,a)=w_{0}(x;\omega ,a)+R_{n}(x;\omega ,a),
\label{RS functions Isot}
\end{equation}%
with%
\begin{equation}
w_{0}(x;\omega ,a)=\frac{\omega }{2}x-\frac{a}{x}  \label{RS functions Isot2}
\end{equation}%
and

\begin{eqnarray}
R_{n}(x;\omega ,a) &=&-\frac{E_{n}\left( \omega \right) }{w_{0}(x;\omega
,a)+w_{n-1}(x;\omega ,a_{1})}  \label{RS functions Isot3} \\
&=&\frac{-2n\omega }{\omega x-\left( 2a+1\right) /x-}\Rsh ...\Rsh \frac{%
2\left( n-j+1\right) \omega }{\omega x-\left( 2\left( a+j\right) -1\right)
/x-}\Rsh ...\Rsh \frac{2\omega }{\omega x-\left( 2\left( a+n\right)
-1\right) /x}.  \notag
\end{eqnarray}

\bigskip As it is well known, the isotonic oscillator eigenstates can be
also expressed in terms of Generalized Laguerre Polynomials (GLP) $\mathit{L}%
_{n}^{\left( \lambda \right) }$ as

\begin{equation}
\psi _{n}\left( x;\omega ,a\right) \sim x^{a}e^{-\omega x^{2}/4}\mathit{L}%
_{n}^{\left( a-1/2\right) }\left( \omega x^{2}/2\right) .  \label{foOI}
\end{equation}

\bigskip This implies that we have%
\begin{equation}
R_{n}(x;\omega ,a)=-\left( \log \left( \mathit{L}_{n}^{\left( a-1/2\right)
}\left( \omega x^{2}/2\right) \right) \right) ^{\prime }=\omega x\frac{%
\mathit{L}_{n-1}^{\left( a+1/2\right) }\left( \omega x^{2}/2\right) }{%
\mathit{L}_{n}^{\left( a-1/2\right) }\left( \omega x^{2}/2\right) },
\label{RSLaguerre}
\end{equation}%
which is singular at the nodes of $\psi _{n}\left( x;\omega ,a\right) $,
that is, at the zeros of $\mathit{L}_{n}^{\left( a-1/2\right) }\left( \xi
\right) $. Concerning these last, we have a classical result of
Kienast,Lawton and Hahn \cite{szego,magnus,erdelyi}:

\emph{Kienast-Lawton-Hahn's Theorem }

Suppose that $\alpha \notin -\mathbb{N}$. Then $\mathit{L}_{n}^{\left(
\alpha \right) }\left( z\right) $ admits

\ \ \ \ \ \ \ 1) $n$ positive zeros if $\alpha >-1$

\ \ \ \ \ \ \ 2) $n+\left[ \alpha \right] +1$ positive zeros if $-n<\alpha
<-1$ ($\left[ \left\vert \alpha \right\vert \right] $ means the integer part
of $\alpha $)

\ \ \ \ \ \ \ 3) No positive zero if $\alpha <-n$

The number of negative zeros is always $0$ or $1$.

\ \ \ \ \ \ \ 1) $0$ if $\alpha >-1$

\ \ \ \ \ \ \ 2) $0$ if $-2k-1<\alpha <-2k$ and $1$ if $-2k<\alpha <-2k+1$,
with $-n<\alpha <-1$

\ \ \ \ \ \ \ 3) $0$ if $n$ is even and $1$ if $n$ is odd, with $\alpha <-n$

This theorem, confirms in particular that, for positive values of $a$, the
RS function $w_{n}(x;\omega ,a)$ corresponding to a physical bound state and
then the associated generalized SUSY partner $V^{\left( n\right) }(x;\omega
,a)$ of $V(x;\omega ,a)$ present always $n$ singularities on the positive
half axis. This family of singular rational extensions of $V(x;\omega ,a)$
will be called the $L0$ series.

\section{\protect\bigskip Confluent hypergeometric equation and isotonic
oscillator}

The confluent hypergeometric equation

\begin{equation}
zy^{\prime \prime }(z;\alpha ,\lambda )+(\alpha +1-z)y^{\prime }(z;\alpha
,\lambda )+\lambda y(z;\alpha ,\lambda )=0  \label{hypergeo}
\end{equation}%
on the positive half real line, can always been reduced to a Schr\"{o}dinger
equation for an isotonic oscillator. Indeed, if we put $z=\omega x^{2}/2$
and $\phi \left( x;\alpha ,\lambda \right) =y(z;\alpha ,\lambda )$ in Eq.(%
\ref{hypergeo}), we obtain the following equation for $\phi \left( x;\alpha
,\lambda \right) $:

\begin{equation}
\phi ^{\prime \prime }(x;\alpha ,\lambda )+(\frac{2\alpha +1}{x}-\omega
x)\phi ^{\prime }(x;\alpha ,\lambda )-2\omega \lambda \phi (x;\alpha
,\lambda )=0
\end{equation}

Then 
\begin{equation}
\psi (x;\alpha ,\lambda )\sim \phi \left( x;\alpha ,\lambda \right) \exp (%
\frac{1}{2}\int dx(\frac{2\alpha +1}{x}-\omega x))=x^{\alpha +1/2}e^{-\omega
x^{2}/4}\phi \left( x;\alpha ,\lambda \right)
\end{equation}
satisfies

\begin{equation}
-\psi ^{\prime \prime }(x;\alpha ,\lambda )+\left( \frac{\omega ^{2}x^{2}}{4}%
+\frac{\left( \alpha +1/2\right) \left( \alpha -1/2\right) }{x^{2}}-\omega
(\alpha +1)\right) \psi (x;\alpha ,\lambda )=2\lambda \omega \psi \left(
x;\alpha ,\lambda \right) .
\end{equation}

If we define $a=\alpha +1/2$, $\psi (x;a-1/2,\lambda )=\psi _{\lambda }(x;a)$
and $E_{\lambda }(\omega )=2\lambda \omega $, we obtain

\begin{equation}
H(\omega ,a)\psi _{\lambda }(x;a)=E_{\lambda }(\omega )\psi _{\lambda }(x;a),
\label{hypergeomod}
\end{equation}%
where 
\begin{equation}
\psi _{\lambda }(x;a)\sim x^{a}e^{-\omega x^{2}/4}y(\omega
x^{2}/2;a-1/2,\lambda ),
\end{equation}%
$H(\omega ,a)$ being the usual isotonic hamiltonian (see Eq.(\ref{isotpot}))

\begin{equation}
H(\omega ,a)=-\frac{d^{2}}{dx^{2}}+V(x;\omega ,a).
\end{equation}

Eq.(\ref{hypergeomod}) is the Schr\"{o}dinger equation \ for the isotonic
oscillator, where for physical bound states we must have $\lambda =n$. In
this case, the confluent hypergeometric equation

\begin{equation}
zy^{\prime \prime }(z;a-1/2,n)+(a+1/2-z)y^{\prime
}(z;a-1/2,n)+ny(z;a-1/2,n)=0,
\end{equation}%
admits the regular solution

\begin{equation}
y(z;a-1/2,n)=L_{n}^{\left( a-1/2\right) }(z)
\end{equation}%
and we have

\begin{equation}
\psi _{n}(x;a)\sim x^{a}e^{-\omega x^{2}/4}L_{n}^{\left( a-1/2\right)
}(\omega x^{2}/2).
\end{equation}

This is exactly the physical state for the isotonic oscillator at the energy 
$E_{n}=2n\omega $.

In fact, as shown by Erdelyi \cite{erdelyi, magnus,gomez,gomez5}, Eq.(\ref%
{hypergeo}) admits quasi rational solutions built from GLP in four sectors
of the values of the parameters $\alpha $ and $\lambda $

\begin{equation}
\left\{ 
\begin{array}{c}
\lambda =n:\ y_{0}(z;\alpha ,n)=L_{n}^{\left( \alpha \right) }(z) \\ 
\lambda =-n-\alpha -1:\ y_{1}(z;\alpha ,\alpha +1+n)=e^{z}L_{n}^{\left(
\alpha \right) }(-z) \\ 
\lambda =n-\alpha :\ y_{2}(z;\alpha ,n-\alpha )=z^{-\alpha }L_{n}^{\left(
-\alpha \right) }(z) \\ 
\lambda =-n-1:\ y_{3}(z;\alpha ,-n-1)=z^{-\alpha }e^{z}L_{n}^{\left( -\alpha
\right) }(-z)%
\end{array}%
\right.
\end{equation}

They correspond to the following four eigenfunctions

\begin{equation}
\left\{ 
\begin{array}{c}
\psi _{n}(x;a)\sim x^{a}e^{-\omega x^{2}/4}L_{n}^{\left( a-1/2\right)
}(\omega x^{2}/2),\quad E_{n}(\omega )=2n\omega \\ 
\psi _{-n-\alpha -1/2}(x;a)\sim x^{a}e^{\omega x^{2}/4}L_{n}^{\left(
a-1/2\right) }(-\omega x^{2}/2),\quad E_{-n-\alpha -1/2}(\omega )=-2\left(
n+\alpha +1/2\right) \omega \\ 
\psi _{n-a+1/2}(x;a)\sim x^{1-a}e^{-\omega x^{2}/4}L_{n}^{-\left(
a-1/2\right) }(\omega x^{2}/2),\quad E_{n-a+1/2}(\omega )=2\left(
n-a+1/2\right) \omega \\ 
\psi _{-n-1}(x;a)\sim x^{1-a}e^{\omega x^{2}/4}L_{n}^{-\left( a-1/2\right)
}(-\omega x^{2}/2),\quad E_{-n-1}(\omega )=-2\left( n+1\right) \omega%
\end{array}%
\right.  \label{secteurs}
\end{equation}

The 3 last cases don't correspond to physical states and physical energies.

\section{Discrete symmetries of the isotonic RS equation}

Since the isotonic oscillator is shape invariant, the $A\left( w_{0}\right) $
DBT is an invariance transformation for the RS\ equations associated to the
family of isotonic oscillators indexed by the couple of parameters $(\omega
,a)$ (see Eq.(\ref{isotpot})). But this family of RS\ equations is covariant
under other specific transformations which act on the parameters of the
isotonic potentials and preserve their functional class. As we will see, the
connections between the quasi rational sectors of the confluent
hypergeometric equation admit a very simple interpretation in terms of
covariance transformations of the isotonic potential.

\subsection{\protect\bigskip Inversion of $\protect\omega $ the parameter}

The first covariance transformation for $V(x;\omega ,a)$ acts on the $\omega 
$ parameter as

\begin{equation}
\omega \overset{\Gamma _{\omega }}{\rightarrow }\left( -\omega \right)
,\left\{ 
\begin{array}{c}
V(x;\omega ,a)\overset{\Gamma _{\omega }}{\rightarrow }V(x;\omega ,a)+\omega
(2a+1) \\ 
w_{n}(x;\omega ,a)\overset{\Gamma _{\omega }}{\rightarrow }v_{n}(x;\omega
,a)=w_{n}(x;-\omega ,a),%
\end{array}%
\right.
\end{equation}%
$v_{n}(x;\omega ,a)$ satisfying ($E_{n}\left( -\omega \right) =-E_{n}\left(
\omega \right) =E_{-n}\left( \omega \right) $)

\begin{equation}
-v_{n}^{\prime }(x;\omega ,a)+v_{n}^{2}(x;\omega ,a)=V(x;\omega
,a)-E_{-\left( n+a+1/2\right) }\left( \omega \right) .  \label{oregpot}
\end{equation}

From Eq.(\ref{RS functions Isot}), Eq.(\ref{RS functions Isot2}) and Eq.(\ref%
{RS functions Isot3}), writing 
\begin{equation}
v_{n}(x;\omega ,a)=v_{0}(x;\omega ,a)+Q_{n}(x;\omega ,a),  \label{oregRSfct}
\end{equation}%
we deduce

\begin{equation}
v_{0}(x;\omega ,a)=-\frac{\omega }{2}x-\frac{a}{x}  \label{oregRSfct2}
\end{equation}%
and

\begin{eqnarray}
Q_{n}(x;\omega ,a) &=&\frac{E_{n}(\omega )}{v_{0}(x;\omega
,a)+v_{n-1}(x;\omega ,a_{1})}  \label{oregRSfct3} \\
&=&-\frac{2n\omega }{\omega x+\left( 2a+1\right) /x+}\Rsh ...\Rsh \frac{%
2\left( n-j+1\right) \omega }{\omega x+\left( 2\left( a+j\right) -1\right)
/x+}\Rsh ...\Rsh \frac{2\omega }{\omega x+\left( 2\left( a+n\right)
-1\right) /x}  \notag \\
&=&-\left( \log \left( \mathit{L}_{n}^{\left( a-1/2\right) }\left( -\omega
x^{2}/2\right) \right) \right) ^{\prime }.  \notag
\end{eqnarray}

Clearly, for $a\geq 1$ $(l\geq 0)$, $v_{n}(x;\omega ,a)$ does not present
any singularity on the positive real half line. This result is coherent with
the above mentioned Kienast-Lawton-Hahn's theorem since the argument of the
GLP $\mathit{L}_{n}^{\left( a-1/2\right) }$ in the expression of $Q_{n}$ is
now a strictly negative value.

Note that we recover exactly the same results if we use the "spatial Wick
rotation" \cite{shnol',samsonov,tkachuk,fellows,grandati2}%
\begin{equation}
w_{n}(x;\omega ,a)\rightarrow v_{n}(x;\omega ,a)=iw_{n}(ix;\omega ,a).
\end{equation}

This means that the $\Gamma _{\omega }$ transformation send the
singularities of $w_{n}$, which are initially all on the real axis, on the
imaginary axis. This explains why the new RS function $v_{n}$ does not
present any singularity on the real line. Finally, comparing Eq.(\ref%
{secteurs}) to Eq.(\ref{oregRSfct}), Eq.(\ref{oregRSfct2}) and Eq.(\ref%
{oregRSfct3}), we see that $\Gamma _{\omega }$ transforms an eigenfunction
of the first sector into an eigenfunction of the second sector and then
coincides with the Kummer's transformation \cite{erdelyi}.

\subsection{Inversion of $a$ the parameter}

The second covariance transformation acts on the $a$ parameter as

\begin{equation}
a\overset{\Gamma _{a}}{\rightarrow }1-a,\left\{ 
\begin{array}{c}
V(x;\omega ,a)\overset{\Gamma _{a}}{\rightarrow }V(x;\omega ,a)+\omega (2a-1)
\\ 
w_{n}(x;\omega ,a)\overset{\Gamma _{a}}{\rightarrow }u_{n}(x;\omega
,a)=w_{n}(x;\omega ,1-a),%
\end{array}%
\right.
\end{equation}%
$u_{n}(x;\omega ,a)$ satisfying

\begin{equation}
-u_{n}^{\prime }(x;\omega ,a)+u_{n}^{2}(x;\omega ,a)=V(x;\omega
,a)-E_{n+1/2-a}\left( \omega \right) .  \label{aregpot}
\end{equation}

From Eq.(\ref{RS functions Isot}), Eq.(\ref{RS functions Isot2}) and Eq.(\ref%
{RS functions Isot3}) we deduce

\begin{equation}
u_{n}(x;\omega ,a)=u_{0}(x;\omega ,a)+P_{n}(x;\omega ,a),  \label{aregRSfct}
\end{equation}%
where%
\begin{equation}
u_{0}(x;\omega ,l)=\frac{\omega }{2}x+\frac{a-1}{x}  \label{aregRSfct2}
\end{equation}%
and

\begin{eqnarray}
P_{n}(x;\omega ,a) &=&\frac{E_{n}(\omega )}{v_{0}(x;\omega
,a)+v_{n-1}(x;\omega ,a_{-1})}  \label{aregRSfct3} \\
&=&\frac{-2n\omega }{\omega x+\left( 2a-3\right) /x-}\Rsh ...\Rsh \frac{%
2\left( n-j+1\right) \omega }{\omega x+\left( 2\left( a-j\right) -1\right)
/x-}\Rsh ...\Rsh \frac{2\omega }{\omega x+\left( 2\left( a-n\right)
-1\right) /x}  \notag \\
&=&-\left( \log \left( \mathit{L}_{n}^{-\left( a-1/2\right) }\left( \omega
x^{2}/2\right) \right) \right) ^{\prime }.  \notag
\end{eqnarray}

If in this case the argument of the GLP in the right hand member is strictly
positive, the associated $\alpha =-\left( a-1/2\right) $ parameter being
strictly negative. In accordance with the Kienast-Lawton-Hahn's theorem, by
taking $a$ sufficently large, we can decrease the number of real zeros and
in particular we can eliminate all the positive zeros. Thus, if $a>n+1/2$, $%
\mathit{L}_{n}^{-\left( a-1/2\right) }\left( \omega x^{2}/2\right) $ is
strictly positive for any value of $x$. This means that $P_{n}(x;\omega ,a)$
and $u_{n}(x;\omega ,a)$ are not singular on $\left] 0,+\infty \right[ $
when $a=n+m+1/2$, with $m>0$.

Note that

\begin{equation}
P_{1}(x;\omega ,a)=\frac{-2\omega }{\omega x+\left( 2a-3\right) /x}%
=Q_{1}(x;\omega ,a-2).
\end{equation}

Finally, comparing Eq.(\ref{secteurs}) to Eq.(\ref{aregRSfct}), Eq.(\ref%
{aregRSfct2}) and Eq.(\ref{aregRSfct3}), we see that $\Gamma _{a}$
transforms an eigenfunction of the first sector into an eigenfunction of the
third sector.

\subsection{Inversion of both parameters $\protect\omega $ and $a$}

Finally, we can also act simultaneously on both parameter as%
\begin{equation}
(\omega ,a)\overset{\Gamma _{a}\circ \Gamma _{\omega }}{\rightarrow }%
(-\omega ,1-a)\left\{ 
\begin{array}{c}
V(x;\omega ,a)\overset{\Gamma _{a}\circ \Gamma _{\omega }}{\rightarrow }%
V(x;\omega ,a)+2\omega \\ 
w_{n}(x;\omega ,a)\overset{\Gamma _{a}\circ \Gamma _{\omega }}{\rightarrow }%
r_{n}(x;\omega ,a)=w_{n}(x;-\omega ,1-a),%
\end{array}%
\right.
\end{equation}%
$r_{n}(x;\omega ,a)$ satisfying%
\begin{equation}
-r_{n}^{\prime }(x;\omega ,a)+r_{n}^{2}(x;\omega ,a)=V(x;-\omega
,1-a)-E_{n}(-\omega )=V(x;\omega ,a)-E_{-\left( n+1\right) }(\omega ).
\label{oaregpot}
\end{equation}

From Eq.(\ref{RS functions Isot}), Eq.(\ref{RS functions Isot2}) and Eq.(\ref%
{RS functions Isot3}) we have

\begin{equation}
r_{n}(x;\omega ,a)=r_{0}(x;\omega ,a)+T_{n}(x;\omega ,a),  \label{oaregRSfct}
\end{equation}%
where%
\begin{equation}
r_{0}(x;\omega ,a)=-\frac{\omega }{2}x+\frac{a-1}{x}=-w_{0}(x;\omega ,a-1)
\label{oaregRSfct2}
\end{equation}%
and

\begin{eqnarray}
T_{n}(x;\omega ,a) &=&\frac{E_{n}(\omega )}{r_{0}(x;\omega
,a)+r_{n-1}(x;\omega ,a_{-1})}  \label{oaregRSfct3} \\
&=&\frac{2n\omega }{\omega x-\left( 2a-3\right) /x+}\Rsh ...\Rsh \frac{%
2\left( n-j+1\right) \omega }{\omega x-\left( 2\left( a-j\right) -1\right)
/x+}\Rsh ...\Rsh \frac{2\omega }{\omega x-\left( 2\left( a-n\right)
-1\right) /x}  \notag \\
&=&-\left( \log \left( \mathit{L}_{n}^{-\left( a-1/2\right) }\left( -\omega
x^{2}/2\right) \right) \right) ^{\prime }.  \notag
\end{eqnarray}

In this case, the argument of the GLP in the right hand member and the
associated $\alpha =-\left( a-1/2\right) $ parameter are both strictly
negative. In accordance with the Kienast-Lawton-Hahn's theorem, by taking $a$
sufficently large, we can have any zero on the negative half line if $n$ is
even ($n=2l$) and one if $n$ is odd. Thus, if $n=2l$ and $a>2l+1/2$, $%
\mathit{L}_{2l}^{-\left( a-1/2\right) }\left( -\omega x^{2}/2\right) $ is
strictly positive for any value of $x$. This means that $T_{2l}(x;\omega ,a)$
and $r_{2l}(x;\omega ,a)$ are not singular on $\left] 0,+\infty \right[ $
when $a=2l+m+1/2$, with $m>0$.

Note that

\begin{equation}
T_{1}(x;\omega ,a)=\frac{-2\omega }{\omega x-\left( 2a-3\right) /x}%
=R_{1}(x;\omega ,a-2).
\end{equation}

Finally, comparing Eq.(\ref{secteurs}) to Eq.(\ref{oaregRSfct}), Eq.(\ref%
{oaregRSfct2}) and Eq.(\ref{oaregRSfct3}), we see that $\Gamma _{a}\circ
\Gamma _{\omega }$ transforms an eigenfunction of the first sector into an
eigenfunction of the fourth sector and corresponds also to a Kummer's
transformation \cite{erdelyi}.

\section{Regular rational extensions of the isotonic oscillator}

Since the transformations considered above are covariance transformations
for the family of isotonic potentials which regularize the RS functions, we
can use these regularized RS\ functions into the finite difference B\"{a}%
cklund algorithm and generate regular isospectral partners for the isotonic
potential.

\subsection{\protect\bigskip Rational extension of the $L1$ series}

$w_{k}$ and $v_{n}$ are associated to the same potential but with different
eigenvalues (cf Eq(\ref{oregpot}))

\begin{equation}
\left\{ 
\begin{array}{c}
-v_{n}^{\prime }(x;\omega ,a)+v_{n}^{2}(x;\omega ,a)=V(x;\omega
,a)-E_{-\left( n+a+1/2\right) }\left( \omega \right) \\ 
-w_{k}^{\prime }(x;\omega ,a)+w_{k}^{2}(x;\omega ,a)=V(x;\omega
,a)-E_{k}\left( \omega \right) ,%
\end{array}%
\right.
\end{equation}%
which means that we can use $v_{n}$ to build a DBT $A\left( v_{n}\right) $
and apply it to $w_{k}$ as

\begin{equation}
w_{k}(x;\omega ,a)\overset{A\left( v_{n}\right) }{\rightarrow }w_{k}^{\left(
n\right) }(x;\omega ,a)=-v_{n}(x;\omega ,a)+\frac{E_{k}(\omega )-E_{-\left(
n+a+1/2\right) }(\omega )}{v_{n}(x;\omega ,a)-w_{k}(x;\omega ,a)},
\label{backL1}
\end{equation}%
where $w_{k}^{\left( n\right) }(x;\omega ,a)$ satisfies

\begin{equation}
-w_{k}^{\left( n\right) \prime }(x;\omega ,a)+\left( w_{k}^{\left( n\right)
}(x;\omega ,a)\right) ^{2}=V^{\left( n\right) }(x;\omega ,a)-E_{k}\left(
\omega \right) ,  \label{oregRSeq}
\end{equation}%
with

\begin{equation}
V^{\left( n\right) }(x;\omega ,a)=V(x;\omega ,a)+2v_{n}^{\prime }(x;\omega
,a).  \label{oregSUSYpart}
\end{equation}

For every $n\geq 0$, $V^{\left( n\right) }(x;\omega ,a)$ is regular on the
positive half line and isospectral to $V(x;\omega ,a)$

\begin{equation}
V^{\left( n\right) }(x;\omega ,a)\underset{iso}{\equiv }V(x;\omega ,a).
\label{oregSUSYpart2}
\end{equation}

Clearly, $w_{-}^{\left( n\right) }(x;\omega ,a)=-v_{n}(x;\omega ,a)$ is also
a solution of Eq(\ref{oregRSeq}) associated to the eigenvalue $E_{-\left(
n+a+1/2\right) }(\omega )<E_{0}\left( \omega \right) =0$. Nevertheless, its
asymptotic behaviour is similar to the one of $w_{-}^{\left( 0\right)
}(x;\omega ,a)=-\omega x/2-a/x$ and consequently

\begin{equation}
\psi _{-}^{\left( n\right) }(x;\omega ,a)\sim \exp \left( -\int
w_{-}^{\left( n\right) }(x;\omega ,a)dx\right)
\end{equation}%
cannot satisfy the boundary condition associated to the physically allowed
eigenstates.

All the physical eigenfunctions of $H^{\left( n\right) }(\omega
,a)=-d^{2}/dx^{2}+V^{\left( n\right) }(x;\omega ,a)$ are then of the form

\begin{equation}
\psi _{k}^{\left( n\right) }(x;\omega ,a)=\frac{1}{\sqrt{E_{k}(\omega
)-E_{-\left( n+a+1/2\right) }(\omega )}}A\left( v_{n}\right) \psi
_{k}(x;\omega ,a),\ k\geq 0
\end{equation}%
and $H^{\left( n\right) }$ is strictly isospectral to $H$.

Since (cf Eq(\ref{oregRSfct2}))%
\begin{equation}
V(x;\omega ,a)+2v_{0}^{\prime }(x;\omega ,a)=V(x;\omega ,a_{1}),
\end{equation}%
Eq(\ref{oregSUSYpart}) and\ Eq(\ref{oregSUSYpart2}) can still be written as

\begin{equation}
V^{\left( n\right) }(x;\omega ,a)=V(x;\omega ,a_{1})+2Q_{n}^{\prime
}(x;\omega ,a)\underset{iso}{\equiv }V(x;\omega ,a).
\end{equation}

For instance, we have for $n=1$

\begin{equation}
V^{\left( 1\right) }(x;\omega ,a)=V(x;\omega ,a_{1})+\frac{4\omega }{\omega
x^{2}+2a+1}-\frac{8\omega \left( 2a+1\right) }{\left( \omega
x^{2}+2a+1\right) ^{2}}
\end{equation}%
and we recover the first rationally-extended radial oscillator obtained by
Quesne \cite{quesne}. For $n=2$, we have immediately from Eq.(\ref{oregRSfct}%
), Eq.(\ref{oregRSfct2}) and Eq.(\ref{oregRSfct3})

\begin{equation}
-v_{2}(x;\omega ,a-1)=\frac{\omega }{2}x+\frac{a-1}{x}+\frac{4\omega x\left(
\omega x^{2}+\left( 2a+1\right) \right) }{\left( \omega x^{2}+\left(
2a+1\right) \right) ^{2}-2\left( 2a+1\right) },
\end{equation}%
which corresponds to the superpotential associated to the second
rationally-extended radial oscillator of the $L1$ series obtained by Quesne 
\cite{quesne}.

More generally, we have

\begin{equation}
Q_{n}(x;\omega ,a)=\left( \log \left( \mathit{L}_{n}^{\left( a-1/2\right)
}\left( -\omega x^{2}/2\right) \right) \right) ^{\prime }.
\end{equation}

In Odake-Sasaki 's approach \cite{odake,odake2,sasaki2}, this corresponds to
a prepotential of the form

\begin{equation}
W_{n}\left( x;\omega ,a\right) =-\frac{\omega }{4}x^{2}+a\log x+\log \left( 
\mathit{L}_{n}^{\left( a-1/2\right) }\left( -\omega x^{2}/2\right) \right)
\end{equation}%
and we recover (up to a shift in $a\rightarrow a+n-2$) the result obtained
in \cite{odake,odake2,sasaki2} and \cite{grandati2} for the potentials
associated to the L1 exceptional orthogonal polynomials.

\subsection{\protect\bigskip Rational extension of the $L2$ series}

As in the preceding case (cf Eq(\ref{aregpot})), we can use $u_{n}$ to build
a DBT $A\left( u_{n}\right) $

\begin{equation}
w_{k}(x;\omega ,a)\overset{A\left( u_{n}\right) }{\rightarrow }w_{k}^{\left(
n\right) }(x;\omega ,a)=-u_{n}(x;\omega ,a)+\frac{E_{k}(\omega
)-E_{n+1/2-a}(\omega )}{u_{n}(x;\omega ,a)-w_{k}(x;\omega ,a)},
\label{backL2}
\end{equation}%
where $w_{k}^{\left( n\right) }(x;\omega ,a)$ satisfies

\begin{equation}
-w_{k}^{\left( n\right) \prime }(x;\omega ,a)+\left( w_{k}^{\left( n\right)
}(x;\omega ,a)\right) ^{2}=U^{\left( n\right) }(x;\omega ,a)-E_{k}(\omega ),
\label{aregRSeq}
\end{equation}%
with

\begin{equation}
U^{\left( n\right) }(x;\omega ,a)=V(x;\omega ,a)+2u_{n}^{\prime }(x;\omega
,a).  \label{aregSUSYpart}
\end{equation}

If $a>n+1/2$, $U^{\left( n\right) }(x;\omega ,a)$ is regular on the positive
half line and isospectral to $V(x;\omega ,a)$

\begin{equation}
U^{\left( n\right) }(x;\omega ,a)\underset{iso}{\equiv }V(x;\omega ,a).
\label{aregSUSYpart2}
\end{equation}

In this case, as for the $L1$ series, we see immediately that $w_{-}^{\left(
n\right) }(x;\omega ,a)=-u_{n}(x;\omega ,a)$ is another solution of Eq(\ref%
{aregRSeq}) associated to the eigenvalue $E_{n+1/2-a}(\omega )(\omega
)<E_{0}\left( \omega \right) =0$. But here again, the asymptotic behaviour
of $w_{-}^{\left( n\right) }(x;\omega ,a)$ is similar to the one of $%
w_{-}^{\left( 0\right) }(x;\omega ,a)=-\omega x/2-(a-1)/x$ and consequently

\begin{equation}
\psi _{-}^{\left( n\right) }(x;\omega ,a)\sim \exp \left( -\int
w_{-}^{\left( n\right) }(x;\omega ,a)dx\right)
\end{equation}%
cannot satisfy the boundary condition associated to the physically
acceptable eigenstates.

All the physical eigenfunctions of $H^{\left( n\right) }(\omega
,a)=-d^{2}/dx^{2}+U^{\left( n\right) }(x;\omega ,a)$ are then of the form

\begin{equation}
\psi _{k}^{\left( n\right) }(x;\omega ,a)=\frac{1}{\sqrt{E_{k}(\omega
)-E_{n+1/2-a}(\omega )}}A\left( u_{n}\right) \psi _{k}(x;\omega ,a),\ k\geq 0
\end{equation}%
and in the $L2$ series, $H^{\left( n\right) }$ is also strictly isospectral
to $H$.

Since (cf Eq.(\ref{aregRSfct2}))%
\begin{equation}
V(x;\omega ,a)+2u_{0}^{\prime }(x;\omega ,a)=V(x;\omega ,a_{-1}),
\end{equation}%
using Eq.(\ref{aregSUSYpart}) and Eq.(\ref{aregSUSYpart2}), we obtain

\begin{equation}
U^{\left( n\right) }(x;\omega ,a)=V(x;\omega ,a_{-1})+2P_{n}^{\prime
}(x;\omega ,a)\underset{iso}{\equiv }V(x;\omega ,a).
\end{equation}

Note that, since $P_{1}(x;\omega ,a)=Q_{1}(x;\omega ,a-2)$, the first
rational extension of this family has the same functional form than the
first rational extension of the preceding family.

For instance, we have for $n=2$%
\begin{equation}
P_{2}(x;\omega ,a)=-\frac{4\omega x\left( \omega x^{2}+\left( 2a-5\right)
\right) }{\left( \omega x^{2}+\left( 2a-5\right) \right) ^{2}+2(2a-5)},
\end{equation}%
which corresponds to Quesne\cite{quesne} second rational extension of the $%
L2 $ series.

We have also, by redefining $a\rightarrow n+a$

\begin{equation}
V(x;\omega ,a_{n-1})\underset{iso}{\equiv }V(x;\omega ,a_{n})+2P_{n}^{\prime
}(x;\omega ,a_{n}),
\end{equation}%
where

\begin{eqnarray}
P_{n}(x;\omega ,a_{n}) &=&-\frac{2n\omega }{\omega x+\left( 2n+2a-3\right)
/x-}\Rsh ...\Rsh \frac{2\left( n-j+1\right) \omega }{\omega x+2\left( \left(
n+a-j\right) -1\right) /x-}\Rsh ...\Rsh \frac{2\omega }{\omega x+\left(
2a-1\right) /x} \\
&=&-\left( \log \left( \mathit{L}_{n}^{-\left( a+n-1/2\right) }\left( \omega
x^{2}/2\right) \right) \right) ^{\prime }  \notag
\end{eqnarray}%
is regular on the positive half line for $a>0$. In Sasaki and al \cite%
{sasaki,sasaki2} formulation, we recover the associated prepotential via

\begin{equation}
W_{n}\left( x;\omega ,a\right) =-\int u_{n}(x;\omega ,a+n)dx=-\frac{\omega }{%
4}x^{2}-\frac{a+n-1}{x}+\log \left( \mathit{L}_{n}^{-\left( a+n-1/2\right)
}\left( \omega x^{2}/2\right) \right)
\end{equation}%
and the family of regular rational extensions obtained is exactly the $L2$
one.

\subsection{Rational extension of the $L3$ series}

Finally, $w_{k}$ and $r_{n}$ being also associated to the same potential but
with different eigenvalues (cf Eq(\ref{oaregpot})), here again we can use $%
r_{n}$ to build a DBT $A\left( r_{n}\right) $ and apply it to $w_{k}$

\begin{equation}
w_{k}(x;\omega ,a)\overset{A\left( r_{n}\right) }{\rightarrow }w_{k}^{\left(
n\right) }(x;\omega ,a)=-r_{n}(x;\omega ,a)+\frac{E_{k}(\omega )-E_{-\left(
n+1\right) }(\omega )}{r_{n}(x;\omega ,a)-w_{k}(x;\omega ,a)},
\label{backL3}
\end{equation}%
where $w_{k}^{\left( n\right) }(x;\omega ,a)$ satisfies

\begin{equation}
-w_{k}^{\left( n\right) \prime }(x;\omega ,a)+\left( w_{k}^{\left( n\right)
}(x;\omega ,a)\right) ^{2}=W^{\left( n\right) }(x;\omega ,a)-E_{k}(\omega ),
\label{oaregRSeq}
\end{equation}%
with

\begin{equation}
W^{\left( n\right) }(x;\omega ,a)=V(x;\omega ,a)+2r_{n}^{\prime }(x;\omega
,a).  \label{oaregSUSYpart}
\end{equation}

If $n=2l$ and $a>2l+1/2$, $W^{\left( 2l\right) }(x;\omega ,a)$ is regular on
the positive half line and isospectral to $V(x;\omega ,a)$

\begin{equation}
W^{\left( 2l\right) }(x;\omega ,a)\underset{iso}{\equiv }V(x;\omega ,a).
\label{oaregSUSYpart2}
\end{equation}

As for the eigenfunctions of $H^{\left( 2l\right) }(\omega
,a)=-d^{2}/dx^{2}+W^{\left( 2l\right) }(x;\omega ,a)$ generated from those
of by the DBT Eq(\ref{backL3}), they are given by

\begin{equation}
\psi _{k}^{\left( 2l\right) }(x;\omega ,a)=\frac{1}{\sqrt{E_{k}(\omega
)-E_{-\left( 2l+1\right) }(\omega )}}A\left( r_{2l}\right) \psi
_{k}(x;\omega ,a),\ k\geq 0
\end{equation}%
and constitute physically allowed eigenstates. But for the $L3$ series the
isospectrality is no more strict as for the preceding series. Indeed, Eq(\ref%
{oaregRSeq}) is evidently satisfied by the regular RS function

\begin{equation}
w_{-}^{\left( 2l\right) }(x;\omega ,a)=-r_{2l}(x;\omega ,a),
\end{equation}%
the asymptotic behaviour of which being identical to the one of $%
w_{-}^{\left( 0\right) }(x;\omega ,a)=-r_{0}(x;\omega ,a)$. Then

\begin{equation}
\psi _{-}^{\left( 0\right) }(x;\omega ,a)=\exp \left( \int dxw_{-}^{\left(
0\right) }(x;\omega ,a)\right) \sim \psi _{0}(x;\omega ,a)
\end{equation}%
and contrarily to the preceding cases

\begin{equation}
\psi _{-}^{\left( 2l\right) }(x;\omega ,a)=\exp \left( \int dxw_{-}^{\left(
2l\right) }(x;\omega ,a)\right)  \label{oaregfond}
\end{equation}%
is a physical state associated to the eigenvalue $E_{-\left( n+1\right)
}(\omega )<0$, that is, the fundamental state of the hamiltonian $H^{\left(
2l\right) }(\omega ,a)$. Consequently, $H^{\left( 2l\right) }$ and $H$ are
only quasi-isospectral in this series, $H^{\left( 2l\right) }$ admitting a
supplementary energy level lower than those of $H$.

Since (cf Eq.(\ref{oaregRSfct2}))%
\begin{equation}
V(x;\omega ,a)+2r_{0}^{\prime }(x;\omega ,a)=V(x;\omega ,a_{-1})-2\omega ,
\end{equation}%
using Eq.(\ref{oaregSUSYpart}) and Eq.(\ref{oaregSUSYpart2}), we obtain

\begin{equation}
W^{\left( n\right) }(x;\omega ,a)=V(x;\omega ,a_{-1})-2\omega
+2P_{n}^{\prime }(x;\omega ,a)\underset{iso}{\equiv }V(x;\omega ,a).
\end{equation}

Since $T_{1}(x;\omega ,a)=R_{1}(x;\omega ,a-2)$, the first rational
extension of this family has the same functional form than the first
rational extension of the $L0$ family.

\ For $n=2$, we have

\begin{equation}
T_{2}(x;\omega ,a)=\frac{-4\omega x\left( \omega x^{2}-\left( 2a-3\right)
\right) }{\left( \omega x^{2}-\left( 2a-3\right) \right) ^{2}+2(2a-3)},
\end{equation}%
which is regular if $a\geq 2$ ($l\geq 1$) and corresponds to Quesne\cite%
{quesne} second rational extension of the $L3$ series.

If we redefine $a\rightarrow 2l+1/2+a$, 
\begin{equation}
T_{2l}(x;\omega ,2l+a+1/2)=-\log \left( \mathit{L}_{2l}^{-\left( a+2l\right)
}\left( -\omega x^{2}/2\right) \right) ^{\prime }
\end{equation}%
and $W^{\left( 2l\right) }(x;\omega ,a+2l+1/2)$ are regular on the positive
half line for $a>0$.

\section{Shape invariance properties of the extensions of the isotonic
oscillator}

As observed initially by Quesne \cite{quesne1,quesne} on the $n=1$ and $n=2$
examples, the rational extended potentials of the $L1$ and $L2$ series
inherit of the shape invariance properties of the isotonic potential.
Several general proofs of this result have been recently proposed \cite%
{odake2,sasaki2}, in particular by Gomez-Ullate et al \cite{gomez5}. In the
present approach, these shape invariance properties can be derived in a very
direct and transparent manner.

\subsection{Shape invariance of the extended potentials of the $L1$ series}

The superpartner of a potential of the $L1$ series $V^{\left( n\right)
}(x;\omega ,a)=V(x;\omega ,a)+2v_{n}^{\prime }(x;\omega ,a)$ is defined as

\begin{equation}
\widetilde{V}^{\left( n\right) }(x;\omega ,a)=V^{\left( n\right) }(x;\omega
,a)+2w_{0}^{\left( n\right) \prime }(x;\omega ,a),\ n\geq 0,
\label{SUSYpartL11}
\end{equation}%
$w_{0}^{\left( n\right) }(x;\omega ,a)$ (see Eq.(\ref{backL1})) being the RS
function associated to the ground level of $V^{\left( n\right) }$ ($%
E_{0}\left( \omega \right) =0$).

We then have

\begin{eqnarray}
\widetilde{V}^{\left( n\right) }(x;\omega ,a) &=&V^{\left( n\right)
}(x;\omega ,a)-2v_{n}^{\prime }(x;\omega ,a)-2\left( \frac{E_{-\left(
n+a+1/2\right) }(\omega )}{v_{n}(x;\omega ,a)-w_{0}(x;\omega ,a)}\right)
^{\prime }  \label{SUSYpartL12} \\
&=&V(x;\omega ,a)-2\left( \frac{E_{-\left( n+a+1/2\right) }(\omega )}{%
v_{n}(x;\omega ,a)-w_{0}(x;\omega ,a)}\right) ^{\prime }.  \notag
\end{eqnarray}

Using Eq(\ref{oregRSfct2}), the shape invariance property of $V(x;\omega ,a)$
in Eq.(\ref{VSIP1}) can also be formulated as%
\begin{equation}
V(x;\omega ,a)+2v_{0}^{\prime }(x;\omega ,a)=V(x;\omega ,a_{1}).
\label{VSIP2}
\end{equation}

Inserting Eq(\ref{VSIP2}) in Eq(\ref{SUSYpartL12}), we obtain

\begin{eqnarray}
\widetilde{V}^{\left( n\right) }(x;\omega ,a) &=&V(x;\omega ,a_{1})-2\left( 
\frac{E_{-\left( n+a+1/2\right) }(\omega )}{v_{n}(x;\omega
,a)-w_{0}(x;\omega ,a)}+v_{0}(x;\omega ,a)\right) ^{\prime }
\label{SUSYpartL13} \\
&=&V^{\left( n\right) }(x;\omega ,a_{1})-2\left( \Delta _{n}^{1}\right)
^{\prime },  \notag
\end{eqnarray}%
where

\begin{equation}
\Delta _{n}^{1}=\frac{E_{-\left( n+a+1/2\right) }(\omega )}{v_{n}(x;\omega
,a)-w_{0}(x;\omega ,a)}+v_{0}(x;\omega ,a)+v_{n}(x;\omega ,a_{1}).
\label{delt1}
\end{equation}

As an example, consider the special case $n=1$. Using Eq(\ref{oregRSfct3}),
we can write

\begin{eqnarray}
\Delta _{1}^{1} &=&-2\omega \left( a+3/2\right) \frac{1}{\frac{E_{1}(\omega )%
}{v_{0}(x;\omega ,a)+v_{0}(x;\omega ,a_{1})}-\omega x}-\omega x-\frac{2a+1}{x%
}+\frac{E_{1}(\omega )}{v_{0}(x;\omega ,a_{1})+v_{0}(x;\omega ,a_{2})} \\
&=&\left( 2a+3\right) \frac{\omega x+\frac{2a+1}{x}}{\omega x^{2}+\left(
2a+3\right) }-\omega x-\frac{2a+1}{x}-\frac{2\omega x}{\omega x^{2}+\left(
2a+3\right) }=-\omega x.  \notag
\end{eqnarray}

We obtain

\begin{equation}
\widetilde{V}^{\left( 1\right) }(x;\omega ,a)=V^{\left( 1\right) }(x;\omega
,a_{1})+2\omega ,
\end{equation}%
which implies that $V^{\left( 1\right) }(x;\omega ,a)$ has the same shape
invariance properties as $V(x;\omega ,a)$.

More generally, using Eq(\ref{oregRSfct3}) and defining $z=-\omega x^{2}/2$
and $\alpha =a+1/2$, we obtain\bigskip 
\begin{eqnarray}
\Delta _{n}^{1} &=&E_{-\left( a+n+1/2\right) }(\omega )\frac{1}{%
Q_{n}(x;\omega ,a)-\omega x}+\left( v_{0}(x;\omega ,a)+v_{0}(x;\omega
,a_{1})\right) +Q_{n}(x;\omega ,a_{1})  \label{delt1n} \\
&=&\frac{2\alpha +2}{x}\frac{\mathit{L}_{n}^{\left( \alpha -1\right) }\left(
z\right) }{\mathit{L}_{n-1}^{\left( \alpha \right) }\left( z\right) +\mathit{%
L}_{n}^{\left( \alpha -1\right) }\left( z\right) }-\omega x\frac{\mathit{L}%
_{n-1}^{\left( \alpha +1\right) }\left( z\right) +\mathit{L}_{n}^{\left(
\alpha \right) }\left( z\right) }{\mathit{L}_{n}^{\left( \alpha \right)
}\left( z\right) }-\frac{2\alpha }{x}.  \notag
\end{eqnarray}

But the generalized Laguerre polynomials satisfy the identity

\begin{equation}
\mathit{L}_{n}^{\left( \alpha \right) }\left( z\right) +\mathit{L}%
_{n-1}^{\left( \alpha +1\right) }\left( z\right) =\mathit{L}_{n}^{\left(
\alpha +1\right) }\left( z\right) ,  \label{recLag1}
\end{equation}%
which gives%
\begin{equation}
\Delta _{n}^{1}=-\omega x\frac{\left( \alpha +n\right) \mathit{L}%
_{n}^{\left( \alpha -1\right) }\left( z\right) +z\mathit{L}_{n}^{\left(
\alpha +1\right) }\left( z\right) -\alpha \mathit{L}_{n}^{\left( \alpha
\right) }\left( z\right) }{z\mathit{L}_{n}^{\left( \alpha \right) }\left(
z\right) }.
\end{equation}

The other fundamental recurrence

\begin{equation}
\left( n+\alpha \right) \mathit{L}_{n-1}^{\left( \alpha \right) }\left(
z\right) -z\mathit{L}_{n}^{\left( \alpha +1\right) }\left( z\right) -\left(
n-z\right) \mathit{L}_{n}^{\left( \alpha \right) }\left( z\right) =0,
\label{recLag2}
\end{equation}%
combined with Eq(\ref{recLag1}) gives then directly

\begin{equation}
\Delta _{n}^{1}=-\omega x,
\end{equation}%
that is,

\begin{equation}
\widetilde{V}^{\left( n\right) }(x;\omega ,a)=V^{\left( n\right) }(x;\omega
,a_{1})+2\omega .  \label{SUSYpartL14}
\end{equation}

Consequently $V^{\left( n\right) }(x;\omega ,a)$ inherits of the shape
invariance properties of $V(x;\omega ,a)$ for every value of $n$.

\subsection{Shape invariance of the extended potentials of the $L2$ series}

The superpartner of a potential $U^{\left( n\right) }(x;\omega
,a)=V(x;\omega ,a)+2u_{n}^{\prime }(x;\omega ,a)$ of the $L2$ series is
defined as

\begin{equation}
\widetilde{U}^{\left( n\right) }(x;\omega ,a)=U^{\left( n\right) }(x;\omega
,a)+2w_{0}^{\left( n\right) \prime }(x;\omega ,a),\ n\geq 0,
\label{SUSYpartL21}
\end{equation}%
$w_{0}^{\left( n\right) }(x;\omega ,a)$ (see Eq.(\ref{backL2})) being the RS
function associated to the ground level of $U^{\left( n\right) }$ . Then

\begin{equation}
\widetilde{U}^{\left( n\right) }(x;\omega ,a)=V(x;\omega ,a)-2\left( \frac{%
E_{n+1/2-a}(\omega )}{u_{n}(x;\omega ,a)-w_{0}(x;\omega ,a)}\right) ^{\prime
}.  \label{SUSYpartL22}
\end{equation}

Using as before, the shape invariance properties of $V(x;\omega ,a)$, this
gives

\begin{eqnarray}
\widetilde{U}^{\left( n\right) }(x;\omega ,a) &=&V(x;\omega ,a_{1})-2\left( 
\frac{E_{n+1/2-a}(\omega )}{u_{n}(x;\omega ,a)-w_{0}(x;\omega ,a)}%
+v_{0}(x;\omega ,a)\right)  \label{SUSYpartL23} \\
&=&U^{\left( n\right) }(x;\omega ,a_{1})-2\left( \Delta _{n}^{2}\right)
^{\prime },  \notag
\end{eqnarray}%
where

\begin{equation}
\Delta _{n}^{2}=E_{n+1/2-a}(\omega )\frac{1}{u_{n}(x;\omega
,a)-w_{0}(x;\omega ,a)}+v_{0}(x;\omega ,a)+u_{n}(x;\omega ,a_{1}).
\label{delt2}
\end{equation}

Using Eq(\ref{oregRSfct3}) and defining $z=\omega x^{2}/2$ and $\alpha
=1/2-a $, this becomes%
\begin{equation}
\Delta _{n}^{2}=\frac{\left( 2n-2a+1\right) \omega }{P_{n}(x;\omega ,a)+%
\frac{2a-1}{x}}+P_{n}(x;\omega ,a_{1})=\omega x\left( \frac{\mathit{L}%
_{n}^{\left( \alpha \right) }\left( z\right) }{\mathit{L}_{n-1}^{\left(
\alpha -1\right) }\left( z\right) }+\frac{(n+\alpha )\mathit{L}_{n}^{\left(
\alpha \right) }\left( z\right) }{-\alpha \mathit{L}_{n}^{\left( \alpha
\right) }\left( z\right) +z\mathit{L}_{n-1}^{\left( \alpha +1\right) }\left(
z\right) }\right) ,  \label{delt2n}
\end{equation}

But the generalized Laguerre polynomials satisfy the identity

\begin{equation}
z\mathit{L}_{n-1}^{\left( \alpha +1\right) }\left( z\right) =(n+\alpha )%
\mathit{L}_{n-1}^{\left( \alpha \right) }\left( z\right) -n\mathit{L}%
_{n}^{\left( \alpha \right) }\left( z\right) ,  \label{recLag4}
\end{equation}%
which combined to Eq(\ref{recLag1}) gives

\begin{equation}
\Delta _{n}^{2}=-\omega x.
\end{equation}

Then $U^{\left( n\right) }(x;\omega ,a)$ has the same shape invariance
properties as $V(x;\omega ,a)$ for every value of $n$, that is

\begin{equation}
\widetilde{U}^{\left( n\right) }(x;\omega ,a)=U^{\left( n\right) }(x;\omega
,a_{1})+2\omega .  \label{SUSYpartL24}
\end{equation}

\subsection{SUSY partners of the $L3$ series extended potentials}

In this case, the superpartner of the extended potential $V^{\left( n\right)
}(x;\omega ,a)=V(x;\omega ,a)+2r_{n}^{\prime }(x;\omega ,a)$ is defined as

\begin{equation}
\widetilde{W}^{\left( n\right) }(x;\omega ,a)=W^{\left( n\right) }(x;\omega
,a)+2\left( -r_{n}^{\prime }(x;\omega ,a)\right) =V(x;\omega ,a),\ n\geq 1
\label{SUSYpartL31}
\end{equation}%
since $-r_{n}(x;\omega ,a)$ is the RS function associated to the ground
level of $W^{\left( n\right) }$.

The SUSY partner of $W^{\left( n\right) }(x;\omega ,a)$ is nothing but the
initial potential $V(x;\omega ,a)$ itself and the DBT $A\left( v_{n}\right) $
is the reciprocal of a SUSY partnership.

\section{Conclusion and perspectives}

In this article, a new method to generate the regular rational extensions of
the isotonic oscillator associated to the $L1$ and $L2$ families of
exceptional Laguerre polynomials is presented. It is based on first order
Darboux-B\"{a}cklund Transformations which are built from excited states RS
functions regularized by using specific symmetries of the isotonic
potential. Starting from this primary shape invariant potential and using
the combination of these symmetries and DBT (as covariance transformations),
we generate four towers of secondary potentials, the four series $L0$, $L1$, 
$L2$ and $L3$. Among them, the potentials belonging to the $L1$ and $L2$
series are regular as well as half of the potentials of the $L3$ series, the
other ones being singular on the positive half line. The secondary
potentials of the $L1$ and $L2$ series inherit of the same translational
shape invariance properties as the primary isotonic potential.

These new potentials being obtained, it is still possible to use the
Krein-Adler theorem \cite{krein,adler} and its subsequent extension obtained
by Samsonov \cite{samsonov2}, to generate other secondary potentials by
applications of some particular $n^{th}$ order DBT.

A similar study can be conducted for the other second category potentials
(Darboux-P\"{o}schl-Teller or Scarf hyperbolic and trigonometric) but also
for the first category potentials. These last \cite{grandati} include the
well known case of the one-dimensional harmonic oscillator \cite%
{shnol',samsonov,tkachuk,gomez,fellows,grandati2} but also the Morse
potential (the regular algebraic deformations of which having already be
obtained by Gomez-Ullate et al \cite{gomez}), the effective radial
Kepler-Coulomb potential and the Rosen-Morse potentials. This work is in
progress and will be the object of a forthcoming paper.

\section{\protect\bigskip Acknowledgments}

I would like to thank A.\ B\'{e}rard, R.\ Milson and C.\ Quesne for
stimulating discussions and very interesting suggestions.

\end{document}